\documentclass[aps,prd, notitlepage, twocolumn,superscriptaddress,floatfix,letterpaper,nofootinbib, longbibliography, 10pt]{revtex4-1}

\usepackage{amsmath, amsthm, amssymb,slashed,mathtools,tabu}
\usepackage{pifont}

\usepackage[usenames, dvipsnames]{color}
\usepackage[svgnames]{xcolor}

\usepackage[colorlinks,citecolor=YaleMidBlue, urlcolor=YaleMidBlue, linkcolor=YaleMidBlue]{hyperref} 
\usepackage{lipsum}
\usepackage{setspace}
\usepackage{multirow}

\usepackage{slashed}
\usepackage[makeroom]{cancel}
\usepackage[normalem]{ulem}
\usepackage{soul}

\usepackage{bm}
\usepackage[normalem]{ulem}
\usepackage{upgreek}
\usepackage{bbm}
\usepackage{sseq}
\usepackage[all,cmtip]{xy}
\usepackage{booktabs}

\usepackage{tikz-cd}
\usepackage{tikz}
\usetikzlibrary{matrix}
\usetikzlibrary{decorations.markings}
\usetikzlibrary{tikzmark,decorations.pathreplacing,positioning}
\usepackage{amsfonts}
\usepackage{datetime}

\usepackage{centernot}
\usepackage{enumitem} 
\usepackage{mathtools,amssymb,varwidth}
\usepackage[utf8]{inputenc}

\usepackage[bbgreekl]{mathbbol}
\usepackage{amscd}

\definecolor{YaleBlue} {HTML}{00356b}
\definecolor{YaleMidBlue} {HTML}{286dc0}
\definecolor{YaleLightBlue} {HTML}{63aaff}

\def\be{\begin{equation}}
\def\ee{\end{equation}}
\newenvironment{myfont}[2][]{\csname#2\endcsname[#1]}{}

\newcommand{\stkout}[1]{\ifmmode\text{\sout{\ensuremath{#1}}}\else\sout{#1}\fi}
\newcommand{\bea}{\begin{eqnarray}}
\newcommand{\eea}{\end{eqnarray}}

\newcommand{\ii}{\hspace{1pt}\mathrm{i}\hspace{1pt}}

\newcommand{\mb}[1]{\mathbf{#1}}

\newcommand{\dd}{\hspace{1pt}\mathrm{d}}

\newcommand{\eq}[1]{(\ref{#1})}

\newcommand{\Tr}{{\rm Tr}}

\newcommand{\bpm}{\begin{pmatrix}}
\newcommand{\epm}{\end{pmatrix}}
\newcommand{\bmm}{\begin{matrix}}
\newcommand{\emm}{\end{matrix}}

\newcommand{\Z}{\mathbb{Z}}

\newcommand{\cB}{ {\cal B} }

\newcommand{\cG}{ {\cal G} }

\newcommand{\SO}{{\rm SO}}
\newcommand{\Spin}{{\rm Spin}}
\newcommand{\U}{{\rm U}}

\newcommand{\rF}{{\rm F}}

\newcommand{\PD}{{\rm PD}}

\newcommand{\nn}{{\nonumber}}

\newcommand{\Fig}[1]{Fig.~\ref{#1}} 

\newcommand{\sRef}[1]{[\onlinecite{#1}]}
\newcommand{\spd}[1]{(#1+1)d}

\def\CB{{\cal B}}

\def\Z{{\mathbb{Z}}}

\def\Tr{{\mathrm{Tr}}}

\def \Hom{\operatorname{Hom}}

\def \H{\operatorname{H}}

\def \Z{\mathbb{Z}}

\def\B{\mathrm{B}}

\begin{document}

\title{Boundary topological orders of (4+1)d fermionic \texorpdfstring{$\Z_{2N}^\rF$}{Z2N} SPT states}

\author{Meng Cheng}
\affiliation{Department of Physics, Yale University, New Haven, CT 06520-8120, USA}

\author{Juven Wang}
\affiliation{London Institute for Mathematical Sciences, Royal Institution, London, W1S 4BS, UK}

\affiliation{Center of Mathematical Sciences and Applications, Harvard University, MA 02138, USA}

\author{Xinping Yang}
\email{xinping.yang@yale.edu}
\thanks{The authors are listed alphabetically.}
\affiliation{Department of Physics, Yale University, New Haven, CT 06520-8120, USA}

\begin{abstract}
     We investigate (3+1)d topological orders in fermionic systems with an anomalous $\mathbb{Z}_{2N}^{\mathrm{F}}$ symmetry, where its $\mathbb{Z}_2^{\mathrm{F}}$ subgroup is the fermion parity. Such an anomalous symmetry arises as the discrete subgroup of the chiral U(1) symmetry of $\nu$ copies of Weyl fermions of the same chirality. Inspired by the crystalline correspondence principle, we deform the anomalous $\Z_{2N}^\mathrm{F}$ symmetry of \spd{3} Weyl fermion to the anomalous $C_N \times \Z_2^\mathrm{F}$ symmetry. Then we microscopically construct symmetry-preserving gapped boundary states of the closely-related (4+1)d $C_N\times \mathbb{Z}_2^{\mathrm{F}}$ symmetry-protected topological (SPT) state (with $C_N$ being the $N$-fold rotation), whenever it is possible. In particular, for $\nu=N$, we show that the (3+1)d symmetric gapped state admits a topological $\mathbb{Z}_4$ gauge theory description at low energy, and propose that a similar theory saturates the corresponding $\mathbb{Z}_{2N}^\mathrm{F}$ anomaly. For $N \nmid \nu$, our construction admits no topological quantum field theory (TQFT) symmetric gapped state; while for $\nu=N/2$, we find a non-TQFT symmetric gapped state via stacking lower-dimensional (2+1)d non-discrete-gauge-theory topological order inhomogeneously. For other values of $\nu$, no symmetric gapped state is possible within our construction, which is consistent with the no-go theorem by Cordova-Ohmori.   
\end{abstract}

\maketitle

\tableofcontents

\section{Introduction}

't Hooft anomaly is a fundamental property of global symmetry in quantum many-body systems and quantum field theories. 
It measures mild violations of gauge invariance in response to external gauge fields, and is an invariant under renormalization group flow 
(i.e. independent of energy scales). This property makes it a powerful tool to constrain the low-energy physics of an interacting system and has found many applications in both condensed matter and high-energy theories. Physically, 't Hooft anomalies can be characterized by an invertible topological field theory (iTFT) in one dimension higher, i.e. a symmetry-protected topological (SPT) phase. This bulk-boundary correspondence has proven instrumental in our understanding of the classification of SPTs.

In general, an anomalous system does not allow a completely featureless ground state. In spacetime dimensions higher than \spd{1}, there are three possible categories of low-energy dynamics: 
\begin{enumerate}
\item a symmetry-preserving gapless state 
e.g. described by a massless free theory or an interacting conformal field theory, 
\item a spontaneous symmetry-breaking state, or 
\item a symmetry-preserving gapped topologically ordered state. 
\end{enumerate}
In the last case, the state must exhibit topological order and can often be described by a topological quantum field theory (TQFT) at low energy. In each case, the anomaly still poses nontrivial constraints on the theories. In particular, various examples have been found in which no TQFTs can be compatible with the anomaly, termed as ``symmetry enforced gaplessness"~\cite{WangSenthilPRB2014, CordovaOhmori1}. 

A well-known example of ``symmetry-enforced gaplessness" 
is the \emph{perturbative local anomaly} of continuous symmetry in \spd{3}; often the continuous symmetry can be chosen to be chiral symmetry of fermions, hence such an anomaly is often called the chiral anomaly~\cite{Adler1969gkABJ, Bell1969tsABJ}.

More recently, examples for \emph{non-perturbative global anomaly}, often of finite group symmetries, have been systematically investigated.\footnote{Perturbative local anomaly is detected by small (i.e. local) continuous symmetry transformation connected to the identity transformation,
hence it is sometimes called the continuous anomaly.

Non-perturbative global anomaly is 
detected by large (i.e. global) discrete symmetry transformation disconnected from the identity transformation,
hence it is sometimes called the discrete anomaly.}
Notably, this was shown to be the case for certain $\Z_N$ chiral anomaly in \spd{3}~\cite{CordovaOhmori2}, which can arise in Weyl fermions with chiral symmetries. Besides their fundamental roles in particle physics, Weyl fermions can also emerge in solid state systems, i.e. the so-called Weyl semimetals~\cite{WSM}. Alternatively, they can be viewed as boundary theories of  \spd{4} fermionic SPT (FSPT) phases.

In this work, we study \spd{3} gapped topological states with $\Z_{2N}^\rF$ chiral anomaly as the boundary theory of \spd{4} $\Z_{2N}^\rF$ FSPT. Suppose the anomaly class is indexed by $\nu$, it can be realized by $\nu$ copies of Weyl fermions $\psi_\alpha$ where $\alpha=1,2,\dots,\nu$, and the symmetry is generated by $\psi_\alpha\rightarrow e^{\ii\pi/N}\psi_\alpha$~\cite{CordovaOhmori2}. The simplest nontrivial example is $N=2$, where the anomaly is classified by 
 the bordism group \cite{2016arXiv160406527F} $\Omega_5^{\Spin \times_{\Z_2^\rF} {\Z_{4}}}=\Z_{16}$~\cite{2018arXiv180502772T, Garc_a_Etxebarria_2019, hsieh2018discretegaugeanomaliesrevisited, 
 Wan_2020, GuoJW1812.11959}, i.e. $\nu$ is defined mod 16.

Interestingly, such a $\Z_{16}$ class shows up as the mixed gauge-gravity non-perturbative global anomaly~\cite{Garc_a_Etxebarria_2019, Wan_2020,
wang2020anomalycobordismconstraintsstandard, wang2020anomalycobordismconstraintsgrand, JW2012.15860,
WangWanYou2112.14765, WangWanYou2204.08393, Putrov_2024} for the Standard Model (SM) with a $\Z_{4,X}$ chiral symmetry, where $X$ is a particular combination of the  baryon minus lepton $\mathbf{B}-\mathbf{L}$ number 
and the properly quantized electroweak hypercharge $\tilde{Y}$, i.e. $X \equiv 5(\mathbf{B}-\mathbf{L})-\frac{2}{3} \tilde{Y}$ proposed by Wilczek-Zee~\cite{KraussWilczekPRLDiscrete1989}.
The $\Z_{4,X}$ symmetry-preserving deformation can 
still violate baryon {\bf B} conservation (which triggers nucleon decays)
or lepton {\bf L} conservation~\cite{PhysRevLett.43.1566, Wilczek1979hcZee}. Note that $X^2=(-1)^\rF$ and $\U(1)_X \supset \Z_{4,X} \supset \Z_2^\rF$, thus $\Z_{4,X} =\Z_{4}^\rF$. The three generations of experimentally confirmed quarks and leptons contribute a total number of Weyl fermions in the SM as $3 \times 15 = 45$, which gives a $\Z_{4, X}$ anomaly index $\nu=45 \mod 16 = -3 \mod 16$ (more details on this anomaly in the SM and the phenomenological implications can be found in Appendix \ref{sec:BSM}). 

We then address the following question:

\begin{quote}
   \textit{What is the minimal topological order saturating the $\Z_{2N}^\rF$ anomaly?} 
\end{quote}

Microscopically, we consider whether the $\Z_{2N}^\rF$ anomaly can be saturated by a symmetric gapped state as the boundary theory of \spd{4} $\Z_{2N}^\rF$ FSPT, realized via the block state construction \cite{SongPRX2017}. Assuming that the symmetric gapped state at the low energy is described by a TQFT, the no-go theorem shown by Cordova and Ohmori in \sRef{CordovaOhmori2} suggests that when $N\nmid \nu$ there exists no gapped boundary TQFT description. On the other hand, when $N\mid \nu$, no such obstructions should exist. Thus we would like to understand what topological orders are sufficient to saturate the anomaly. In particular, we provide a microscopic construction of the symmetric gapped boundary state, which can be unfolded in the following steps:

\begin{enumerate}
	\item  We start from  \spd{3} Weyl fermions. They have anomalous $\Z_{2N}^\rF$ symmetry and can be viewed as the boundary states of a \spd{4} $\Z_{2N}^{\rF}$ FSPT.
    In Section \ref{sec:crystalline}, we consider a UV deformation of the Weyl fermions by a spatially-inhomogeneous mass term. It breaks the $\Z_{2N}^\rF$ symmetry while preserving a $C_N\times\Z_2^\rF$ symmetry. The resulting $C_N\times \Z_2^\rF$ symmetry is still anomalous, which manifests as chiral fermions appearing on the \spd{1} rotational axis (while the fermions are gapped away from the axis). The number of chiral fermions signaling the anomalous $C_N \times \Z_2^\rF$ symmetry is precisely given by the anomaly index $\nu$ of $\Z_{2N}^\rF$ symmetry of the Weyl fermion. In fact, these two symmetry groups, $\Z_{2N}^\rF$ and $C_N\times \Z_2^\rF$, are known to be related via the ``crystalline correspondence principle"~\cite{ThorngrenElse2018,  Cheng:2018aaz, ZhangPRB2020, Manjunath:2022hbm}. Namely, the classifications of invertible TQFTs with these symmetries are isomorphic. For SPT phases, the classification turns out to be $\Z_{16}$ for both cases when $N=2$.
	\item To gap out the chiral fermions on the \spd{1} rotational axis, in Section \ref{sec:block state} we generalize the approach in [\onlinecite{Yang2024}] to provide explicit constructions of gapped boundary states for $\nu$ being a multiple of $N$. Namely, the boundary microscopic construction inherits from the bulk block state by taking a spatial slice with additional invertible phases $\cal{F}$'s arranged in a $C_N$-symmetric way. To symmetrically gap out the gapless modes, we tentatively extend the $C_N$ to a larger symmetry group $C_{N'}$ to trivialize the $C_N$ anomaly
	\[
	1 \rightarrow G_A \rightarrow C_{N'} \rightarrow C_N \rightarrow 1\,.
	\]
	Meanwhile, the invertible phases $\cal{F}$'s are enriched by $G_A$. Then we gauge $G_A$ to restore the physical symmetry $C_N$ and to convert the enriched $\cal{F}$'s into invertible defects in the boundary $G_A$ gauge theory. 
	
	Within our construction, we give a slight refinement of Cordova-Ohmori's no-go theorem:
	\begin{itemize}
		\item [--] for $\nu = N$, i.e. FSPTs characterized by Majorana chain decoration, the \textit{minimal} topological order saturating the $\Z_{2N}^\rF$ anomaly is a $\Z_4$ gauge theory; 
		\item [--] for $\nu = \frac{N}{2}$, i.e. FSPTs characterized by $p_x + \ii p_y$ decoration, there exists no gapped boundary TQFT description but a highly-anisotropic symmetric gapped boundary state. We conjecture that same is true for boundary states of other \spd{4} FSPTs with $p_x + \ii p_y$ decoration;
		\item [--] for other values of $N$  our construction shows no symmetric gapped boundary states, consistent with the no-go theorem, and we conjecture that the boundary state must be gapless or $C_N$ symmetry breaking.
	\end{itemize}
	
\end{enumerate}

\subsection{\texorpdfstring{$\Z_{2N}^\rF$}{Z2N} symmetry and the nonperturbative global anomaly}

The anomalous $\Z_{2N}^\rF$ symmetry can be realized as a discrete chiral symmetry of Weyl fermion. For a left-handed (or right-handed) Weyl fermion $\psi$ in \spd{3}, the chiral U(1) symmetry is defined as $\psi\rightarrow e^{\ii\theta}\psi$. It is well-known that the U(1) symmetry has a 't Hooft anomaly, which can be characterized via the anomaly-inflow mechanism by a Chern-Simons term $A \dd A \dd A$ of the background U(1) gauge field $A$ in the (4+1)d bulk. This anomaly is perturbative and the low-energy dynamics must be gapless if the symmetry is not explicitly broken. We are interested in breaking the symmetry group to $\Z_{2N}^\rF$. The generator $g$ of $\Z_{2N}^\rF$ symmetry acts on the Weyl fermion as 
\begin{equation*}
    \psi\rightarrow e^{\frac{\ii \pi }{N}}\psi.
\end{equation*}
Note that $g^N$ gives the fermion parity $\Z_{2}^\rF$, hence the notation $\Z_{2N}^\rF$. In the relativistic theory, the full spacetime and internal symmetry of Weyl fermion is
\begin{equation*}
    \text{Spin}(3,1)\times_{\Z_2^\rF}\Z_{2N}.
\end{equation*}

The $\Z_{2N}^\rF$ symmetry is still anomalous. Again by anomaly inflow, $\Z_{2N}^\rF$ anomalies can be classified by $\Z_{2N}^\rF$ fermionic SPT phases in (4+1)d. We are mostly interested in $N=2^p$. In this case, the Freed-Hopkins classification~\cite{2016arXiv160406527F} 
of \spd{4} $\Z_{2N}^\rF$ SPTs is given by the 5d bordism group $\Omega_5$ of 
$$
{\rm Spin}^{\Z_{2N}} \equiv \Spin \times_{\Z_2^\rF} {\Z_{2N}} \equiv
\frac{\Spin \times {\Z_{2N}}}{\Z_2^\rF}$$ 
symmetry
~\cite{hsieh2018discretegaugeanomaliesrevisited, GuoJW1812.11959}: 
\footnote{{Note that for a product of abelian groups, 
for clarity, we use $\times$ for the symmetry groups,
while the direct sum $\oplus$ for the group classification of SPT states.}}
\begin{equation*}
    \Omega_5^{{\rm Spin}^{\Z_{2^{p+1}}}}=\Z_{2^{p+3}}\oplus \Z_{2^{p-1}},
\end{equation*}
which matches our calculation of the $C_N \times \Z_2^\rF$ FSPT classification using block state construction in Appendix \ref{App:C_N classification}. For general $N$, we can always write $N= 2^p \cdot q$ where $q$ is a generic odd integer, and the symmetry group factorizes as 
$
\Z_{2N}^\rF =\Z_{2^{p+1}}^\rF\times \Z_q$. In this case, the classification turns out to be 
\bea
\Omega_5^{{\rm Spin}^{\Z_{2^{p+1}  \cdot q}}} &=&
\Omega_5^{{\rm Spin}^{\Z_{2^{p+1}}}}\oplus \Omega_5^{\rm Spin}(B\Z_q)\cr
&\cong&
\Omega_5^{{\rm Spin}^{\Z_{2^{p+1}}}}\oplus \tilde{\Omega}_5^{\rm SO}(B\Z_q)
\eea
where the latter group is 
$\Omega_5^{\rm Spin}(B\Z_q) \cong \tilde{\Omega}_5^{\rm SO}(B\Z_q)$
isomorphic to 
the bosonic SPT phases protected by $\Z_q$ symmetry. 
Here $\tilde{\Omega}_5^{\rm SO}(\B G) \equiv\Omega_5^{\rm SO}(\B G)/\Omega_5^{\rm SO}$ is the reduced bordism group, modding out the $\Omega_5^{\rm SO}=\Omega_5^{\rm SO}(pt)$.
Moreover, if we write
$N= 2^p \cdot q = 
2^p \cdot 3^r \cdot s$ where $s$ is a generic odd integer
that contains no factor of 3, then
\bea
&&\Omega_5^{{\rm Spin}^{\Z_{2^{p+1}  \cdot  3^r \cdot s
}}} \cong
\Omega_5^{{\rm Spin}^{\Z_{2^{p+1} 
}}}\oplus \tilde\Omega_5^{\rm SO}(B\Z_{3^r \cdot s}  )
 \cr
&&=
\Z_{2^{p+3}}\oplus \Z_{2^{p-1}}\oplus
\Z_{3^{r+1}}\oplus \Z_{3^{r-1}}\oplus 
\Z_s \oplus 
\Z_s. \nn
\eea
TQFTs on the boundary of bosonic SPT phases, especially the ``beyond-group-cohomology" ones, have been systematically studied in [\onlinecite{Yang2024}]. Hence, in this work, we focus on the $N=2^p$ case, where the fermionic part (from $\Z_{2^{p+1}}^\rF$ symmetry) plays a role.

However, not all of the \spd{4} $\Z_{2N}^\rF$ SPT phases admit a gapped boundary TQFT description. To be concrete, denote the anomaly index by $\nu$, which can be realized by $\nu$ copies of Weyl fermions. In [\onlinecite{C_rdova_2020}], Cordova and Ohmori showed that when $N\nmid \nu$, a unitary spin TQFT cannot exist as the boundary theory to saturate the anomaly discussed above. We now briefly review their argument. Consider the SPT theory on $M_4 \times \mathbb{R}_+$, where $M_4$ is a 4D spin manifold\footnote{We denote $n$D for space dimension,
and $(n+1)d$ for spacetime dimension.} with Pontryagin number 48, and $\mathbb{R}_+$ is the upper half line $\mathbb{R}_+ = \{x^5 | x^5 \geqslant 0 \}$. Here the nontrivial SPT phase lives in the space $ x^5 > 0 $, while the trivial phase lives in $ x^5 <0$. The boundary theory $\mathcal{B}$ lives on $x^5 = 0$. Let $Z_{\mathcal{B}}[M_4]$ be the path integral of $\mathcal{B}$ on the interval $0 \leqslant x^5 \leqslant x$, $x > 0$, which defines a boundary state $|\mathcal{\mathcal{B}} \rangle$ on $M_4 \times \{x\}$. Their analysis shows that if $\mathcal{B}$ is a symmetry-preserving boundary theory, then when $N\nmid \nu$ we must have 
\[
Z_{\mathcal{B}}[M_4] = 0.
\]
This condition is satisfied when $\mathcal{B}$ is the \spd{3} massless Weyl fermion because of fermion zero modes. When $\mathcal{B}$ is a unitary spin TQFT $\mathcal{B}_{\mathrm{top}}$, consider the theory $\mathcal{B}_{\mathrm{top}}$ on a simply-connected spin manifold $X_4$.
One can then show that the partition function  $Z_{\mathcal{B}_{\mathrm{top}}}(X_4)$ is always non-zero, by relating $|Z_{\mathcal{B}_{\mathrm{top}}}(X_4)|^2$ to $Z_{\mathcal{B}_{\mathrm{top}}}(S^4)$ and $Z_{\mathcal{B}_{\mathrm{top}}}(S^2\times S^2)$, and applying the result from \sRef{cordova2020anomalyobstructionssymmetrypreserving} to show that $Z_{\mathcal{B}_{\mathrm{top}}}(S^2\times S^2) \neq 0$. Taking $X_4$ to have Pontryagin number 48, we find a contradiction. In conclusion, a gapped boundary TQFT does not exist for \spd{4} $\Z_{2N}^\rF$ SPTs when $N\nmid \nu$.

\subsection{\texorpdfstring{$C_N \times \Z_2^\rF$}{CN} symmetry}\label{sec:crystalline}
In this section, we discuss the UV deformation from the anomalous $\Z_{2N}^\rF$ symmetry to the anomalous $C_N \times \Z_2^\rF$ symmetry.
Starting from the (3+1)d Weyl fermion, we turn on an ``s-wave" superconducting pairing, which completely gaps out the Weyl fermion. More explicitly, the superconducting Weyl fermion is described by the following Hamiltonian:
\begin{equation*}
    H=\int \dd^3\mb{x} \left[\psi^\dag(\mb{x}) 
 \bm{\sigma}\cdot \mb{p}\psi(\mb{x}) + \Delta_0 \psi^\dag(\mb{x}) \ii \sigma_y \psi^\dag(\mb{x})+\text{h.c.}\right].
\end{equation*}
The pairing breaks $\Z_{2N}^\rF$ to $\Z_2^\rF$, but preserves the full spatial rotation group SO(3), under which $\psi$ transforms as spin-1/2 (since it is a Lorentz spinor). 

However, instead of the uniform pairing $\Delta_0$, let us also introduce a (1+1)d vortex line passing through the origin. More explicitly, in cylindrical coordinates ($\rho=\sqrt{x^2+y^2}$, $\theta=\arctan\frac{y}{x}$, $z$), we have $\Delta(\rho, \theta, z)=|\Delta_0| f(\rho) e^{\ii \theta}$, where $f(\rho)$ is the profile of the order parameter, which approaches $1$ as $\rho\rightarrow \infty$, and vanishes as $\rho\rightarrow 0$. 
It is well-known that the vortex line traps a \spd{1} Majorana-Weyl fermion with a chiral central charge $c_-=1/2$. 

With the non-uniform pairing, the internal $\U(1)$ symmetry and Spin(2) rotation symmetry are broken down to their diagonal subgroup SO(2). To see this, naively, the presence of the vortex line breaks the Spin(2) rotation symmetry. However, the SO(2) subgroup symmetry can be restored, if a spatial rotation $R(\alpha)$ around $z$ by an angle $\alpha$ is combined with a U(1) transformation: 
\begin{equation*}
    \psi(\mb{x})\rightarrow e^{\frac{\ii}{2}\alpha}R(\alpha)\psi(\mb{x}).
\end{equation*}
This way, the additional phase of the superconducting order parameter under rotation is canceled. Because of the $\alpha/2$ phase, the symmetry group is changed from Spin(2) to SO(2).

More formally, without the pairing the full spacetime and internal symmetry of Weyl fermion is the Lorentz group with the chiral U(1) symmetry
$$\Spin(3,1) \times_{\Z_2^\rF} \U(1),$$
which becomes 
$$\Spin(1,1) \times_{\Z_2^\rF} \U(1)$$ on the gapless vortex line. And the rotational symmetry around the vortex line is 
$$\Spin(2) \times_{\Z_2^\rF} \U(1) \supset \SO(2).$$

To summarize, by introducing a \spd{1} superconducting vortex line, 
the \spd{3} Weyl fermion is deformed into a \spd{1} Majorana-Weyl fermion along the vortex line while everywhere else is gapped. This deformation preserves a diagonal subgroup of Spin(2) and U(1), which is just the SO(2) rotation. The presence of the chiral modes along the vortex line indicates that under the unbroken SO(2) symmetry, the theory is still anomalous.

We will consider further breaking SO(2) down to $C_N$ (e.g. by the lattice), which provides an example of $C_N\times \Z_2^\rF$ anomaly.

This example can be viewed as a special case of the fermionic crystalline correspondence principle~\cite{Cheng:2018aaz,ZhangPRB2020,Manjunath:2022hbm}. 
In this case, the principle gives an isomorphism between the SPT phases protected by $\Z_{2N}^\rF$ symmetry, and those protected by $C_N\times \Z_2^\rF$ symmetry. For related discussions of the boundary TQFT correspondence between SPTs protected by internal symmetries and those protected by crystalline symmetries, see
[\onlinecite{zhang2021crystallineequivalentboundarybulkcorrespondence}, \onlinecite{zhang2023anomalousboundarycorrespondencetopological}].

The example of a Weyl fermion deformed by a superconducting vortex can also be viewed as the boundary of a (4+1)d $C_N\times \Z_2^\rF$ FSPT state. 
To be more explicit, let us parameterize the four Euclidean coordinates of the (4+1)d FSPT bulk as $(x,y,z,w)$ and the rotational plane along the $x$-$y$ plane is parameterized by $(x,y)$.
In four spatial dimensions, the rotational center (that is invariant and fixed under the $x$-$y$ plane rotation) is also a 2D plane as the $z$-$w$ plane. A ``fixed-point" state for (4+1)d $C_N\times \Z_2$ FSPT phase is to simply put a (2+1)d $p_x+\ii p_y$ superconductor on the rotation center $z$-$w$ plane. This is an example of the so-called ``block-state" construction~\cite{SongPRX2017}. On the \spd{3} boundary, the 
(1+1)d Majorana-Weyl fermion on the (1+1)d rotation axis (i.e. the vortex line) is precisely the (1+1)d edge mode of the (2+1)d $p_x+\ii p_y$ superconductor.  

More generally, one can start from $\nu$ copies of Weyl fermions and follow the same procedure to arrive at $\nu$ Majorana-Weyl fermions at the rotation axis. Below, we say that this system has $C_N\times \Z_2^\rF$ anomaly indexed by $\nu$. Alternatively, one may view such a system as the boundary of a \spd{4} $C_N \times \Z_2^\rF$ SPT state, which takes the form of a stack of $\nu$ copies of $p_x+\ii p_y$ superconductors at the 
(2+1)d rotation center. In Appendix \ref{App:C_N classification}, we derive the full classification of $C_N\times \Z_2^\rF$ FSPTs in (4+1)d using the block-state construction.

\section{Construction of symmetric gapped boundary} \label{sec:block state}

\begin{figure*}
    \centering
    \includegraphics[width=\textwidth]{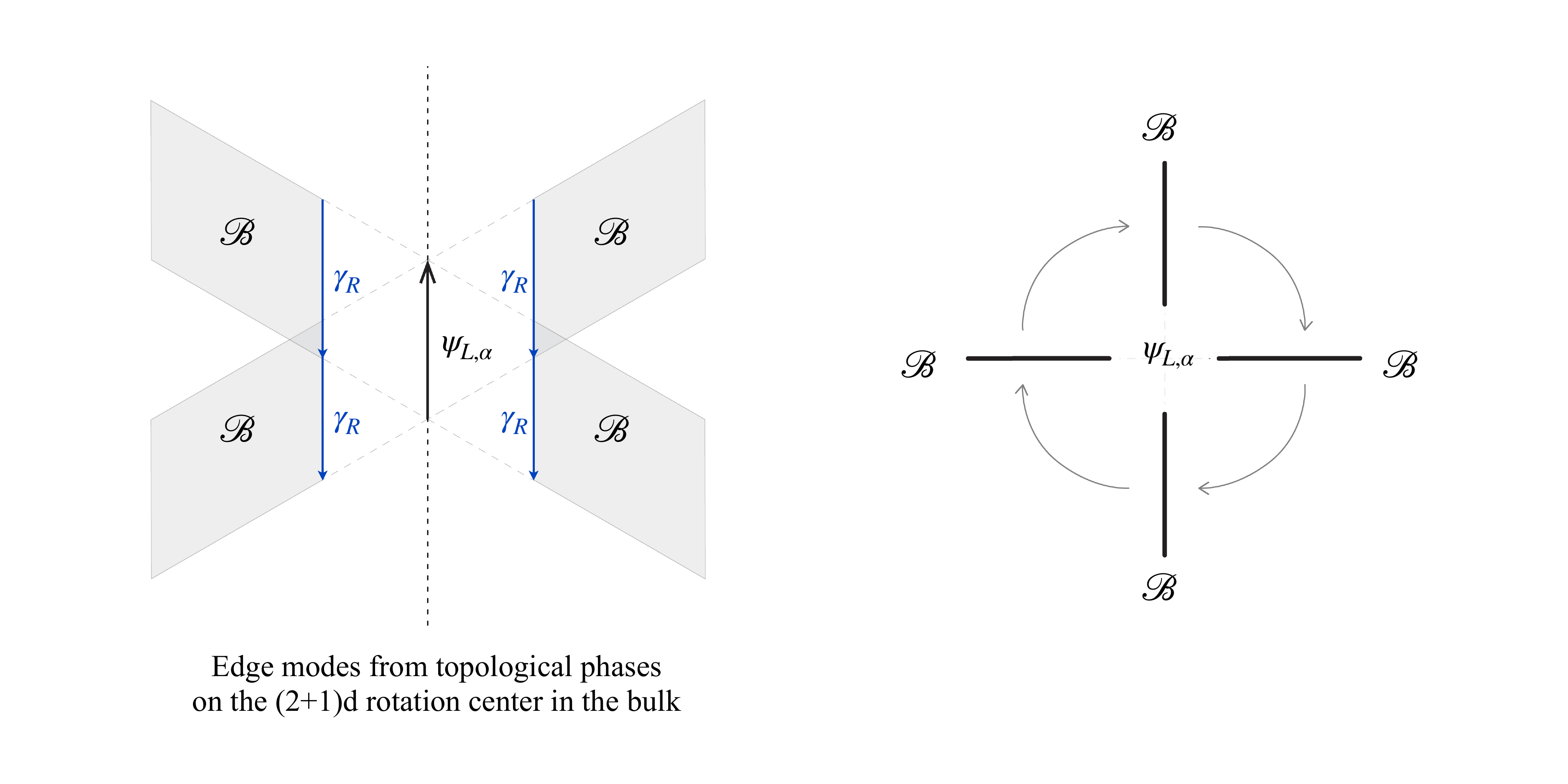}
    \caption{Block construction for a \spd{3} symmetric gapped state on the boundary of \spd{4} bulk $C_N \times \Z_2^\rF$ FSPT state. Here we demonstrate the $N=4$ case.}
    \label{fig:Z4_1}
\end{figure*}

 The (3+1)d deformed theory has $\nu$ copies of (1+1)d Majorana-Weyl fermions on the rotation axis (i.e. the $z$ axis) and each of them carries a chiral central charge $c_-=1/2$. 
 In order to create a fully gapped \spd{3} boundary, we follow the strategy in \cite{Yang2024} and use the construction shown in Fig.\ref{fig:Z4_1}. Here the 3D $x$-$y$-$z$ boundary space is parametrized by $(x,y,z)$, taken to be the $w=0$ slice of the 4D $x$-$y$-$z$-$w$ space of (4+1)d bulk. Choose $N$ half planes all terminating at the $z$ axis, the positions of which are related to each other by $C_N$ rotation. 
 For example, one of the half-planes could be the plane $y=0, x\geq 0$, all the others are obtained by $C_N$ rotations.

 On each \spd{2} plane we decorate a \spd{2} (chiral) topological order $\cal{B}$, again all placed in a $C_N$-symmetric way. 
 At the 1D rotational center $z$-axis, we have $N$ of the (1+1)d edge modes from the (2+1)d topological phases $\mathcal{B}$'s 
 on the half-planes and the (1+1)d edge conformal field theory (CFT) from the 2D rotation center (namely the $z$-$w$ plane) in the \spd{4} bulk. 
 We require that these edge modes together can be gapped out while preserving the $C_N$ symmetry. 

We will distinguish two types of gapped boundary states: TQFT and non-TQFT. A TQFT boundary state is one that can be described by a TQFT at low energy. In \spd{3}, a full classification of TQFTs in bosonic systems (or non-spin) has been achieved recently~\cite{Lan4dTQFT1, Lan4dTQFT2, Johnson-Freyd:2020usu}. Roughly speaking, they are all finite group gauge theories. In other words, they are all topologically equivalent to gauged (fermionic or bosonic) SPT states.  {We believe the same is true for fermionic theories, as one can always gauge the fermion parity and apply the bosonic classification.}

The block construction, naively, does not lead to \spd{3} TQFT boundary states since the insertion of \spd{2} (chiral) topological orders breaks the Lorentz symmetry. However, if $\CB$ is a \spd{2} topological gauge theory, then one can ``deform" the block construction to a TQFT by converting $\cB$'s into \emph{invertible topological defects} in a \spd{3} TQFT. We fill the boundary with a \spd{3} topological gauge theory and then condense the bound state of 2D and 3D gauge charges such that there are no deconfined anyons on the \spd{2} planes. 

For simplicity, it suffices to assume that the gauge group is an Abelian group $G_A$ for all our examples. 
Then $\CB$ has a set of bosonic or fermionic anyons carrying gauge charges of $G_A$. 
They are labeled by characters (i.e. one-dimensional representations) of $G_A$, 
the collection of which forms the group $\hat{G}_A := \rm{Hom}(G_A,\U(1))$, 
the Pontryagin dual of $G_A$. 
Importantly, these anyons braid trivially with each other, so they form a Lagrangian subgroup. 
Here being a Lagrangian subgroup means that every anyon NOT in $\hat{G}_A$ must braid nontrivially with at least one anyon in $\hat{G}_A$. 
We denote anyons from \spd{2} topological gauge theory by $b_{2d}$ for $b\in \hat{G}_A$. In addition, they may transform nontrivially under $C_N$, e.g. $C_N$ may act projectively on the anyons.

Then on the \spd{3} boundary, we stack a \spd{3} topological gauge theory with the same gauge group $G_A$ and the same group of anyons $\hat{G}_A$. Importantly, we require that as Abelian anyon theories, the \spd{3} theory $\hat{G_A}$ must be identical to the \spd{2} theory $\CB$, meaning that we choose 3D gauge charges $b_{3d}$ that share the same transformation properties as $b_{2d}$ under $C_N$. On each $\mathcal{B}$ plane, we condense the bound state $b_{2d}\bar{b}_{3d}$ for all $b\in \hat{G_A}$, which effectively identifies $b_{2d}$ with $b_{3d}$. By construction, the condensation preserves the $C_N$ symmetry and it is expected that there is no spontaneous symmetry breaking. 
After the condensation, all the \spd{2} planes become codimension-1 invertible topological defects 
in the \spd{3} topological gauge theory, implementing the 0-form $C_N$ symmetry. Below we denote the invertible topological defect by $\mathsf{D}$. 

To see why $\mathsf{D}$ is invertible, it is useful to consider an alternative description of the construction, adopting the ``symmetry-extension" method in [\onlinecite{Wang2017locWWW1705.06728}]. First, we symmetrically gap out the gapless modes by introducing invertible phases $\cal{F}$ in the block-state construction. To do this, the global symmetry $C_N$ may need to be extended by the ``gauge group" $G_A$ in nontrivial ways, and consequently $\cal{F}$ is enriched by the extended symmetry group. Then to restore the physical symmetry $C_N$, one gauges the symmetry group $G_A$ in the entire \spd{3} boundary space to produce a $G_A$ gauge theory. The invertible state $\cal{F}$ becomes an invertible defect $\mathsf{D}$ in the gauged theory.~\footnote{Before gauging, $\cal{F}$ being invertible means that there exists another state ${\cal{F}}^{-1}$, such that the stacking of $\cal{F}$ and ${\cal F}^{-1}$ can be smoothly deformed to a completely trivial state without breaking the protecting symmetry. Importantly, this deformation can be generated by a local symmetry-preserving Hamiltonian. This implies that after gauging, the defects corresponding to $\cal{F}$ and ${\cal{F}}^{-1}$ are also inverse of each other.} If instead we only gauge the symmetry $G_A$ on the \spd{2} states, they turn into topological phases described by $\cal{B}$ in the previous construction. As we will see in Section \ref{sec:boundary construction}, the computations are easier to handle in this ``ungauged" description.

We want to emphasize that the block-state construction is the UV description of the boundary state. In the IR limit, the $C_N$ symmetry in the block state construction should become the product of the continuous rotation symmetry and the internal ``emanant" symmetry of the IR theory. More concretely, denote by $r_{\rm UV}$ the generator of the $C_N$ rotation group. Then we have the identification  $r_{\mathrm{UV}} = U \times r_{\mathrm{IR}}(\frac{2 \pi}{N})$ where $U$ is the internal symmetry.  In the TQFT, the IR rotation symmetry $r_{\mathrm{IR}}$ does not act and $r_{\mathrm{UV}}$ becomes the internal $\Z_N$ symmetry in the TQFT, implemented by the invertible topological defect $\mathsf{D}$.

In the following gapped boundary constructions, we only need to focus on that $N$ is a power of 2 (namely $N=2^p$), for the reasons discussed in the introduction;
other cases are related to the bosonic cases already explored in \cite{Yang2024}. Without loss of generality, assume that the anomaly index is $\nu\geq 0$, which means there are $\nu$ Majorana-Weyl fermions on the rotation axis (namely, the $z$ axis on the (3+1)d boundary).

\subsection{General considerations}

 The topological order $\CB$ must have $c_-=-\frac{\nu}{2N}$, so it must be non-Abelian unless $\nu$ is a multiple of $N$. Since our goal is to create a fully gapped state, it must be that $N$ copies of the topological order can have a gapped boundary to an invertible state, even before adding symmetry considerations. 
 Mathematically, it means that $\CB$ has order $N$ in the ``fermionic Witt group"~\cite{davydov2011structurewittgroupbraided, DGNOWitt1}. Witt group is the equivalence classes of topological orders, with the following equivalence relation: $\CB_1$ and $\CB_2$ are equivalent, if and only if $\CB_1$ and $\CB_2$ can have a gapped interface up to the stacking of invertible states. The structure of the Witt group of fermionic topological orders is highly constrained. In particular, any non-Abelian class in the fermionic Witt group has order either 2 or infinity~\cite{DGNOWitt1}.
  On the other hand, from the chiral central charge, one can easily show that the order of $\CB$ is a multiple of $\frac{N}{(N,\nu)}$, {where $(N,\nu)$ is the greatest common divisor of $N$ and $\nu$}. We thus require $\frac{N}{(N,\nu)}=2$, or $\nu=\frac{N}{2}$. If $\nu\neq 0$ or $N/2$ (mod $N$), the construction cannot work. We conjecture that in these cases, the boundary must be gapless or $C_N$ symmetry-breaking.

\subsection{\texorpdfstring{$\nu=\frac{N}{2}$}{nu}: non-TQFT gapped state}
\label{sec:non-TQFT}

Let us start from $N=2$ and $\nu=\frac{N}{2}=1$. We fix an arbitrary plane passing through the rotation axis. The $C_2$ rotation reduces to the reflection on this plane. 
Exactly the same problem was studied for the surface of the crystalline topological superconductor~\cite{Fidkowski1305.5851, cheng2018microscopic1707.02079}, and one can construct a (2+1)d $\mathrm{SO}(3)_3\boxtimes \, \text{semion-fermion}$ topological order (TO) preserving all symmetries~\cite{cheng2018microscopic1707.02079}. Recall that surface anomalies of \spd{3} crystalline topological superconductors are classified by a $\Z_{16}$ index $k$. 
The $\mathrm{SO}(3)_3$ Chern-Simons theory\footnote{{Note that the
$\mathrm{SO}(3)_3$ Chern-Simons theory can be constructed out of the anyon contents of 
$\mathrm{SU}(2)_6$ Chern-Simons theory but only keeping its integer-spin representation 
($j=0,1,2,3$) of
SU(2) while removing the half-integer-spin representation ($j=1/2,3/2,5/2$)
of SU(2). It should not be confused with the Fibonacci anyon theory, which can be described as
the subcategory of integer-spin representations ($j=0,1$)
of $\mathrm{SU}(2)_3$ Chern-Simons theory. In our convention, Fibonacci anyon theory is roughly the 
``$\mathrm{SO}(3)_{\frac{3}{2}}$'' Chern-Simons theory.}} 
fermionic TO carries the anomaly index $k=3$~\cite{Wang_2017, cheng2018microscopic1707.02079, MaoPRR2020}, and the semion-fermion TO theory carries the anomaly index $-2$ (by choosing the appropriate symmetry fractionalization on the semion)~\cite{Fidkowski1305.5851}. Thus together they have anomaly index $k=3+(-2)=1$. By the dimensional reduction to the plane, it is clear that $k$ and $\nu$ are in fact equal.

However, the $\mathrm{SO}(3)_3$ fermionic TO is not a topological {discrete} gauge theory, and cannot be made so by stacking with TOs. This is because anyons in a topological gauge theory must have quantum dimensions square to an integer,~\footnote{In fact, either $\Z$ or $\Z\sqrt{2}$ where the $\sqrt{2}$ is from Majorana zero modes.} but the spin-1 anyon in ${\rm SO(3)}_3$ TO has quantum dimension $2+\sqrt{2}$. Our previous construction of turning the non-TQFT state into a TQFT state does not apply. The boundary state we have constructed is therefore non-TQFT.

For general {even integer $N \in 2 \mathbb{Z}$}, one can easily adopt the $N=2$ construction. Namely, one views the system as $N/2$ identical copies of the $N=2$ boundary. 
{That is, the $n$-th copy is rotated from the $0$-th copy (namely the $N$-th copy on the $x$-$z$ plane)
by a $\frac{2\pi n}{N}$ angle, spanning the $x$-$y$ plane.}

\subsection{\texorpdfstring{$\nu=N$: $\Z_4$}{Z4} gauge theory}\label{sec:boundary construction}

\begin{figure*}
    \centering
    \includegraphics[width=\textwidth]{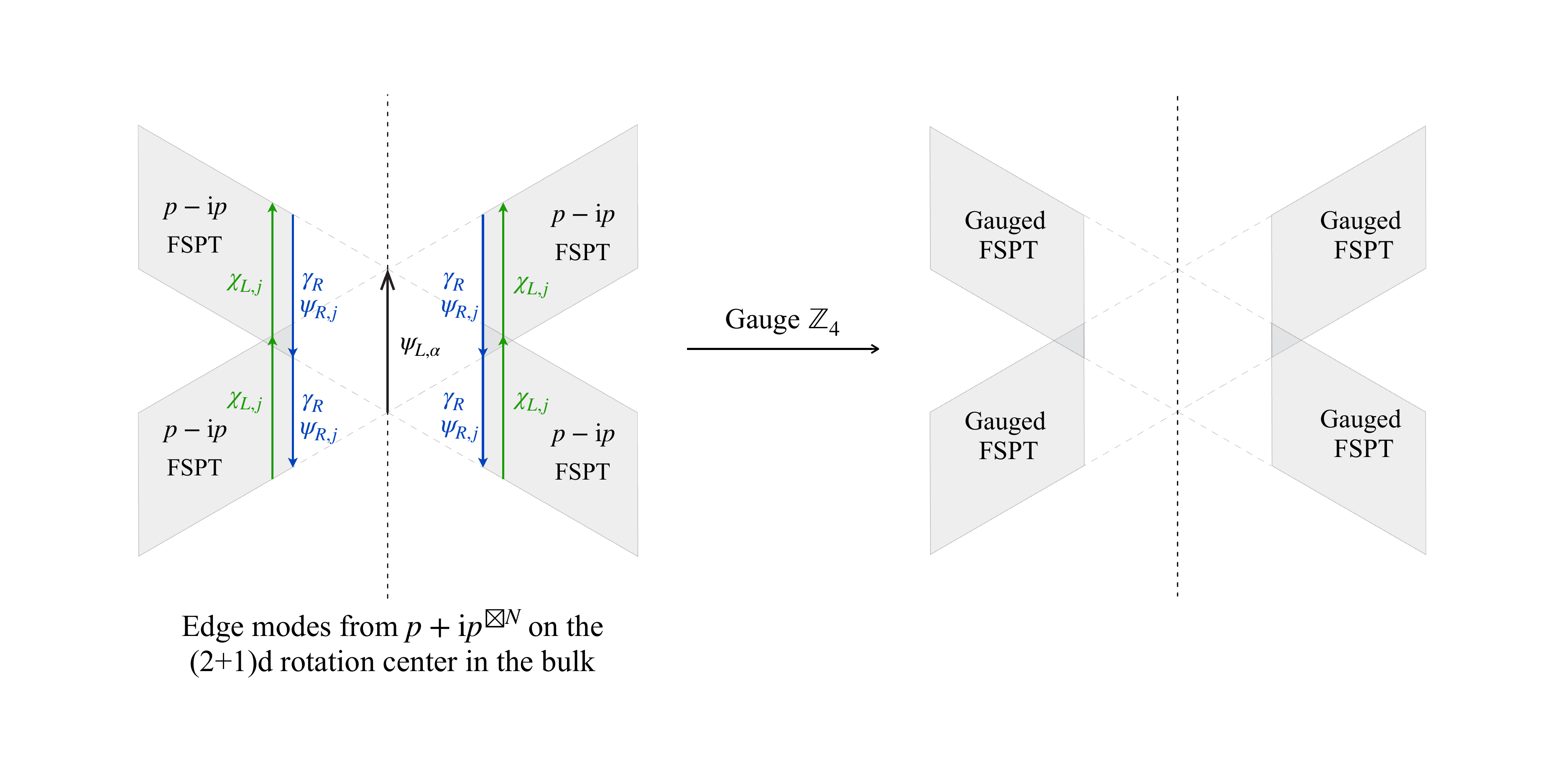}
    \caption{The left figure demonstrates the ``ungauged" description of \spd{3} $N=4$  symmetric gapped boundary construction: $p_x - \ii p_y$ superconductors are added to cancel the chiral central charge, then we further decorate the \spd{2} planes with $\Z_4 \times \Z_2^\rF$ SPTs to gap out everything. The right figure shows the actual TQFT with a static network of gauged FSPT defects after we gauge the $\Z_4$ symmetry.}
    \label{fig:Z4}
\end{figure*}    

Recall the block-state construction discussed in Section~\ref{sec:crystalline}, \spd{4} $C_N\times \Z_2^\rF$ SPT state takes the form of $\nu$ copies of $p_x+\ii p_y$ superconductors at the \spd{2} rotation center in the bulk, giving rise to $\nu$ chiral edge modes on the \spd{1} axis on the boundary. To construct a gapped boundary for $\nu=N$ class FSPT, we first need to cancel the chiral central charge. This can be achieved by using the trivial block state consisting of $N$ layers of $p_x- \ii p_y$ superconductors~\cite{SongPRX2017, Cheng:2018aaz}, where the $C_N$ symmetry acts as the $\Z_N$ cyclic permutation on the layers. Each \spd{2} $p_x-\ii p_y$ superconductor contributes a \spd{1} Majorana-Weyl field (often called chiral Majorana fermion field in condensed-matter literature) $\gamma_{R,j}, j=1,\dots, N$. Under the $\Z_N$ generator, it transforms as
\begin{equation*}
    \gamma_{R,j}\rightarrow \gamma_{R,j+1}.
\end{equation*}
Let us define 
\begin{equation}
    \psi_{R,l} = \frac{1}{\sqrt{N}}\sum_j \omega^{-lj}\,\gamma_{R,j},\;\, \omega=e^{\frac{2\pi \ii}{N}}.
\end{equation}
{This definition of the basis change converts the $\Z_N$ symmetry generator from the shift matrix form to the diagonal clock matrix form.} In this new basis, the chiral fermion fields transform under the $\Z_N$ symmetry as 
\begin{equation*}
\psi_{R,l}\rightarrow \omega^l\psi_{R,l},
\end{equation*}
satisfying $\psi_{R,N-l}=\psi_{R,l}^\dag$. Thus we have $\psi_{R,\alpha},\, \alpha=1,\dots, {N/2-1}$ as independent Weyl fields, and $\psi_{R,0}$, $\psi_{R,N/2}$ as Majorana-Weyl fields. To distinguish them from Weyl fields, we rename these two as $\chi_{R,0}$ and $\chi_{R,N/2}$.

As mentioned earlier, the 1D rotation axis hosts $N$ chiral edge modes from $N$ copies of $p_x + \ii p_y$. Similarly, after Fourier transform, there are $N$ Majorana-Weyl fermions: $\gamma_{L,j}$, $j=1,2,\dots, N$. Living on the rotation axis, these chiral fields are all invariant under $C_N$ rotation symmetry and thus can be combined into $N/2$ Weyl fermion: $\psi_{L,\alpha},\,\alpha=1,\cdots, {N/2}$.  We can choose to couple $\psi_{L,N/2}$ with ${\chi}_{R,0}$ via a mass term, e.g.
\begin{equation*}
    (\psi_{L,N/2}-\psi^\dag_{L,N/2})\,\chi_{R,0}
\end{equation*}
 leaving a single Majorana-Weyl fermion $\chi_{L,0}$ defined as: 
 \begin{equation}
     \chi_{L,0}=\psi_{L,N/2}+\psi_{L,N/2}^\dag.
 \end{equation}
To simplify notations, from now on we rename $\chi_{L,0}$ as $\chi_L$, ${\chi}_{R,N/2}$ as ${\chi}_R$.

Denote the generator of the $C_N$ symmetry by $r$. The field content of the \spd{1} theory is listed in the following table:
\begin{center}
\begin{tabular}{|c|c|c|}
 \hline
    Field & Index & $r$ action\\
    \hline
     $\psi_{L,l}$ & $l=1,\cdots, \frac{N}{2}-1$ & $1$\\[1ex] 
     ${\psi}_{R,l}$ & $l=1,\cdots, \frac{N}{2}-1$ & $e^{\frac{2\pi \ii}{N}l}$ \\[1ex] 
     $\chi_L$ & & $1$ \\[1ex] 
     ${\chi_R}$ & & $-1$\\[1ex] 
     \hline
\end{tabular}
\end{center}
Note that the left-moving edge modes from the $p_x+\ii p_y$'s are at the rotation center, thus do not transform under the $r$ symmetry; while the right-moving edge modes from the $p_x-\ii p_y$'s are placed around the rotation center, thus do transform under the $r$ symmetry.

In order to gap out these modes, we generalize the ``symmetry extension" method ~\cite{Witten2016cio1605.02391, Wang2017locWWW1705.06728} to our case. The idea is to enlarge the symmetry group so that the SPT order becomes trivialized, but then gauge the extra symmetries to recover the physical symmetry group. In this way, one obtains a topological gauge theory as the boundary state. To determine the candidate gauge group, we give the following intuition: The smallest gauge group is $\Z_2$. Suppose we extend $C_N$ by $\Z_2$ (i.e. ungauging the $\Z_2$ symmetry), the \spd{2} planes carry $\Z_2 \times \Z_2^\rF$ SPTs. Knowing that \spd{2} $\Z_2 \times \Z_2^\rF$ SPTs are $\Omega_3^{\rm Spin}(B\Z_2) \cong \Z_8$ classified and any of the so-called Majorana SPTs here is of order 8. The $\Z_2$ gauge group cannot work unless $N$ is a multiple of 8. So for general $N = 2^p$, the next candidate gauge group is $\Z_4$ (while $\Z_3$ is ruled out for obvious reasons). In the following calculations, we first check if a $\Z_4$ gauge theory is enough to saturate the anomaly for $N = 2^p$, then for 8 $| \,N$, we check whether a $\Z_2$ gauge theory could be the minimal TQFT to saturate the anomaly (minimal in the sense of being the smallest gauge group).

\subsubsection{\texorpdfstring{$N=2^p$}{N=2p}}

In the first case, we extend $C_N$ symmetry by $\Z_4$ and we find that the following group extension: 
   \begin{equation*}
       1\rightarrow \Z_4\rightarrow C_{4N}\rightarrow C_N\rightarrow 1.
   \end{equation*}
Namely, the gauge group is $\Z_4$ and the symmetry group $C_N$ is extended to $C_{4N}$.
Denote the generator of $\Z_4$ by $g$.  More explicitly, the group extension means $r^N=g$. 

As illustrated in Fig.\ref{fig:Z4}, upon ungauging the $\Z_4$ symmetry, the \spd{2} planes are further dressed with \spd{2} Majorana FSPTs protected by $\Z_4\times\Z_2^{\rF}$. Recall that the classification of $\Z_4\times\Z_2^\rF$ FSPT phases in \spd{2} is {$\Omega_3^{\rm Spin}(B\Z_4) \cong\Z_8\oplus \Z_2$} ~\cite{ChengFSPT2018, WangFSPT2017, GuoJW1812.11959}. 

In particular, the Majorana FSPT is the generator of the $\Z_2$ classification subgroup.
This phase has the \spd{1} edge theory consisting of four Majorana-Weyl fermions $\chi_{L,j}, \, j=1,\cdots, 4$, and 
two Weyl fermions ${\psi}_{R,j},\, j=1,2$. Under the $\Z_4$ symmetry action $g$, they transform as:
\begin{equation*}
 g:\chi_{L,1}\rightarrow -\chi_{L,1}, \;\,{\psi}_{R,j}\rightarrow \ii {\psi}_{R,j}.   
\end{equation*}
{Other fields ($\chi_{L,j}, \, j=2,3,4$) are invariant under the $g$ symmetry transformation.}
Most notably, in the Majorana FSPT phase a $g$ symmetry twist defect carries a Majorana zero mode, which is evident from the action on $\chi_{L,1}$. 

Now we create $N$ copies of the Majorana FSPTs placed around the rotation axis in a $C_N$-symmetric way. Their edge modes are denoted by $\chi_{L,j,\alpha}$ and ${\psi}_{R,j,\alpha}$. {Here we use a triplet to label the indices: $L$ or $R$ indicates the chirality of the edge mode, $j$ specifies different edge modes in a given layer $\alpha$ and $\alpha = 1,\cdots,N$ is the layer index.} We postulate the following $r$ symmetry:
\begin{alignat*}{2}
    &\chi_{L,1,\alpha}\rightarrow \chi_{L,1,\alpha+1}, \quad  && 1\leq \alpha\leq N-1.\\
    &\chi_{L,1, N}\rightarrow - \chi_{L,1, 1}, \quad && \\
    &\chi_{L,j,\alpha}\rightarrow \chi_{L,j,\alpha+1}, \quad &&\alpha =1,2, \dots, N, \; j=2,3,4. \\
    &\psi_{R,j,\alpha} \rightarrow e^{\frac{2\pi \ii}{4N}} \psi_{R,j,\alpha+1}, \quad && \alpha =1,2, \dots, N, \; j=1,2,3,4.
\end{alignat*}
Note that the phase factors are chosen so that $r^N=g$ is satisfied. 

As before, we Fourier transform the chiral fermion modes:
\begin{equation*}
\begin{alignedat}{2}
   \psi_{L,1,l}&\sim \sum_{\alpha=1}^N e^{-\frac{\pi \ii}{N}l\alpha}\,\chi_{L,1,\alpha}, \quad && l=1,3,\cdots, N-1.\\
    \psi_{L,j,n}&\sim \sum_{\alpha=1}^N e^{-\frac{2\pi \ii}{N}n\alpha}\,\chi_{L,j,\alpha}, \quad && n=0,1,\cdots, N-1,\; j>1.\\
    {\psi}_{R,j,n} &\sim \sum_{\alpha=1}^N e^{-\frac{2\pi \ii}{N}n\alpha}\,{\psi}_{R,j,\alpha}, \quad && n=0,1,\cdots, N-1. 
\end{alignedat}    
\end{equation*}

Again notice that for $j>1$, $\psi_{L,j,0}$ and $\psi_{L,j,\frac{N}{2}}$ are Majorana-Weyl fermions, while $\psi_{L,j,n}^\dag = \psi_{L,j,N-n}$, giving us $\frac{N}{2}-1$ independent Weyl fermions. For $j=1$, all $\psi_{L,1,l}$ are Weyl fermions. Rename $\psi_{L,j,0},\, \psi_{L,j,\frac{N}{2}}$ as ~$\eta_{L,j,0}$ and $\eta_{L,j,\frac{N}{2}}$ to emphasize that they are Majorana-Weyl fermions, we summarize the field content in the following table:

\begin{center}
    \begin{tabular}{|c|c|c|}
    \hline
         Field & $r$ action & Index range \\
         \hline
         $\psi_{L,1,l}$ & $e^{\frac{\pi \ii}{N}l}$ & $l=1,3,\dots,N-1$\\ [1ex] 
         $\psi_{L,j,l}$ & $e^{\frac{2\pi \ii}{N}l}$ & $l=1,2,\dots, \frac{N}{2}-1$ and $j=2,3,4$\\[1ex] 
         $\eta_{L,j,0}$ & $1$ & $j=2,3,4$\\[1ex] 
         $\eta_{L,j,\frac{N}{2}}$ & $-1$ & $j=2,3,4$\\[1.2 ex] 
         ${\psi}_{R,j,n}$ & $e^{\frac{2\pi \ii}{4N}}e^{\frac{2\pi \ii}{N}n}$ & $n=0,1,\dots, N-1$ and $j=1,2$\\[1ex] 
         \hline
    \end{tabular} 
\end{center}

We want to check whether all of the \spd{1} edge modes together can be gapped out without breaking the $C_{4N}$ symmetry (which becomes a $\Z_{4N}$ symmetry on the \spd{1} modes). This is equivalent to asking whether the symmetry has no 't Hooft anomaly. To this end,  we consider the topological spin $h_r$ of the $r$ twist field, which corresponds to a $2\pi$ self-rotation of the $r$ twist defect in the \spd{2} bulk SPT state. Or equivalently, it is the phase factor from the modular $\mathcal{T}$ transformation on the 2-torus partition function with the $r$ twist field background:
\begin{equation}
   \theta_r=\exp( 2 \pi \ii h_r). 
   \end{equation} 
  This phase factor determines whether the corresponding SPT phase is trivial or not. If it cannot be trivialized to 1 by attaching local excitations, such as symmetry charges or a physical fermions,  to the twist defect, then the SPT phase is nontrivial. In other words, $\Theta_r=\theta_r^{4N}$ is in fact a topological invariant for the $\Z_{4N}$ SPT phase. If $\Theta_r\neq 1$, the \spd{1} edge modes cannot be gapped out without breaking the protecting symmetry. When $\Theta_r=1$, as reviewed in Appendix \ref{App:C_N classification}, there is one more invariant to check, namely whether the twist defect carries a Majorana zero mode (MZM) or not. If both invariants vanish, it should be possible to gap them out with appropriate local interactions while preserving the symmetries, although we will not attempt to determine the gapping terms explicitly. Notice that the \spd{1} fermion modes come in two groups. Those on the rotation center are from $N$ of the $p_x-\ii p_y$ layers and $N$ of the $p_x+\ii p_y$ layers, their contribution to $h_r$ is given by
\begin{equation}
    h_r^{(1)} = -\frac12\sum_{l=1}^{\frac{N}{2}-1} \Big(\frac{l}{N}\Big)^2-\frac{1}{16}.
    \label{eqn:hr1}
\end{equation}
The rest of the contribution comes from the Majorana FSPT layers and is given by
\begin{equation}
\begin{aligned}
    h_r^{(2)} = &\frac{1}{2} \sum_{\stackrel{l=1}{l\text{ odd}}}^{N-1}\left(\frac{l}{2N}\right)^2  + \frac{3}{2} \sum_{l=1}^{\frac{N}{2}-1} \left(\frac{l}{N}\right)^2\\ 
    &+ 3\times\frac{1}{16} - \sum_{n=0}^{N-1} \left(\frac{n}{N}+\frac{1}{4N}\right)^2 \,.
\end{aligned}
\end{equation}
Together we have $h_r=h_r^{(1)}+h_r^{(2)}$:
\begin{equation}
    h_r=h_r^{(1)}+h_r^{(2)} = \frac{-13N^2+12N+4}{48N},
\end{equation}
thus the invariant for $\Z_{4N}$ anomaly in this \spd{1} theory is given by 
\begin{equation}
 \begin{aligned}
    \Theta_r&=e^{2\pi \ii (4N h_r)}\\
    &=\exp \Big[\frac{\ii \pi }{6} \Big(-13N^2+12N+4 \Big) \Big]\\
    &= \exp \Big[-\frac{\ii \pi }{6} \Big(N^2-4 \Big) \Big]\,.
\end{aligned}   
\end{equation}

The $N=2$ case gives $\Theta_r=1$. Then for $N=2^p$ with $p\geq 2$, we have
\begin{equation}
    \frac{N^2-4}{6}=\frac{2^{2p}-4}{6}=\frac{2(2^{2p-2}-1)}{3}.
\end{equation}
Since $2p-2$ is even, one can show that $2^{2p-2}-1$ is a multiple of $3$, using
\begin{equation*}
    x^{2k}-1=(x+1)(x^{2k-1}-x^{2k-2}+\cdots -1).
\end{equation*}
Therefore $\Theta_r=1$ as well. 

Next, we check the second invariant which determines if the $r$-twist defect carries a MZM. We count the number of Majorana fields transforming non-trivially under $r$, and there are four such Majorana fields:
$\chi_{R,0}$ and $\chi_{L,\frac{N}{2},j}$ for $j=2,3,4$. Thus the $r$ twist does not have any MZMs. 

Therefore, the $\Z_{4N}$ symmetry is non-anomalous and all the edge modes can be fully gapped out without breaking the $\Z_{4N}$ symmetry. After this we gauge the $\Z_4$ symmetry to obtain a \spd{3} TQFT. Note that if we only gauge the symmetry in the \spd{2} Majorana FSPT state,  the unit $\Z_4$ gauge flux becomes a non-Abelian anyon with quantum dimension $\sqrt{2}$.~\footnote{In fact, the topological order of the gauged Majorana FSPT can be identified as $\text{T-Pfaffian}\boxtimes \text{semion-fermion}$.} In conclusion, we have constructed a \spd{3} $\Z_4$ gauge theory that saturates the $\nu=N$ anomaly.

We can examine how the $C_N$ rotation symmetry acts on various objects in the \spd{3} $\Z_4$ gauge theory. The low-energy excitations include $\Z_4$ gauge charges and $\Z_4$ fluxes. The rotation symmetry acts projectively on gauge charges, meaning that a unit charge picks up a projective phase $e^{\frac{\ii \pi}{2N}}$ under an $r$-action. Physically, this can be viewed as the ``fractionalized" angular momentum. The rotational action on $\Z_4$ fluxes is facilitated by the codimension-1 gauged $\Z_4\times \Z_2^\rF$ FSPT defects. In the presence of a unit disinclination line, which can be viewed as a flux of the $C_N$ symmetry, a $\Z_4$ flux loop linked to it carries a MZM as it intersects with the gauged FSPT defect.

\subsubsection{\texorpdfstring{$N=0$}{N=0} mod 8}

For $N\neq 0$ (mod 8), we argued that $\Z_4$ gauge theory is the minimal TQFT. 
However, when $8\mid N$, a more careful analysis is needed to determine the minimal TQFT. Below we carry out the computations to check if a $\Z_2$ gauge theory is enough to saturate the anomaly when $N$ is a multiple of 8. The analysis is completely parallel with that of the $\Z_4$ gauge group considered above, so we will keep it brief.

As before, we introduce the trivial configuration of $N$ layers of $p_x-\ii p_y$ superconductors, and the topological spin of the $r$ twist field is again given in \eqref{eqn:hr1}. Ungauging the $\Z_2$ symmetry, We then extend the symmetry to $C_{2N}$ via the following short exact sequence:
 \begin{equation*}
       1\rightarrow \Z_2\rightarrow C_{2N}\rightarrow C_N\rightarrow 1.
\end{equation*}
Denote the generator of $\mathbb{Z}_2$ symmetry as $g$, the group extension suggests $r^N = g$. As before, $\mathbb{Z}_2$ would eventually become the gauge symmetry. The \spd{2} planes around the rotational axis are further decorated by $\mathbb{Z}_2 \times \mathbb{Z}_2^\rF$ Majorana FSPT. The generating phase has the edges modes of one left-moving Majorana-Weyl fermion $\chi_L$ and one right-moving Majorana-Weyl fermion $\chi_R$. Under the $g$ symmetry, the edge modes transform as
\begin{equation*}
    g: \chi_L \rightarrow -\chi_L,\;\, \chi_R \rightarrow \chi_R.
\end{equation*}
For $N$ copies of the Majorana FSPTs, we denote the edge modes by $\chi_{\alpha,L}$, $\chi_{\alpha,R}$, and they transform under the $r$ symmetry as
\begin{equation*}
    \begin{split}
    &\chi_{\alpha,L} \rightarrow \chi_{\alpha+1,L}\quad \alpha= 1, \cdots, N-1\,,\\
    &\chi_{N,L} \rightarrow -\chi_{N,L}\,,\\
    &\chi_{\alpha,R} \rightarrow \chi_{\alpha+1,R}\,.
    \end{split}   
\end{equation*}
We perform the Fourier transform
\begin{equation}
\begin{alignedat}{2}
    &\psi_{l,L}\sim \sum_{\alpha=1}^N e^{-\frac{\pi \ii}{N}l\alpha}\,\chi_{\alpha,L} &&\quad l=1,3,\cdots, N-1\,.\\
    &\psi_{n,R}\sim \sum_{\alpha=1}^N e^{-\frac{2\pi \ii}{N}n\alpha}\,\chi_{\alpha,j} &&\quad n=0,1,\cdots, N-1 \,.
\end{alignedat}
\end{equation}
The topological spin of the $r$ twisted fields is
\begin{equation}
  \begin{aligned}
    h_r^{(2)} = \frac{1}{2}\sum_{\stackrel{l=1}{l\text{ odd}}}^{N-1}\left(\frac{l}{2N}\right)^2  - \frac{1}{2} \sum_{l=1}^{\frac{N}{2}-1} \left(\frac{l}{N}\right)^2 -\frac{1}{16} \,.
\end{aligned}  
\end{equation}

Suppose we have $n$ copies of the fundamental $\mathbb{Z}_2 \times \mathbb{Z}_2^\rF$ Majorana FSPTs, the invariant for the $\mathbb{Z}_{2N}$ anomaly is
\begin{equation}
  \begin{aligned}
    \Theta_r &=e^{2\pi \ii 2N (h_r^{(1)}+nh_r^{(2)})}\\&= \exp\left(- \frac{\ii \pi }{12}  \big(N^2+3n+2\big)\right)\,.
\end{aligned}  
\end{equation}

Since $N^2+3n+2$ is odd for even $N$ and odd $n$, we see that $\Theta_r\neq 1$. From there, we conclude that a $\Z_4$ gauge theory is the minimal TQFT to saturate the $\nu=N$ anomaly.

\section{Conclusion and discussions}

In this work, we systematically studied \spd{3} gapped phases with anomalous $\Z_{2N}^\rF$ symmetry. The anomaly is classified by an index $\nu$, and can be realized in $\nu$ left-handed Weyl fermions as the $\Z_{2N}$ subgroup of the global U(1) symmetry. The Cordova-Ohmori constraint forbids any symmetry-preserving TQFTs with such an anomaly when $N\nmid \nu$, but leaves the $N\mid \nu$ case open. To explicitly construct a symmetry-preserving fully gapped state for $\nu=N$, we utilize crystalline correspondence principle to relate $\Z_{2N}^\rF$ and $C_N\times \Z_2^\rF$. We first deform the Weyl fermions by a superconducting mass term containing a vortex line, which changes the symmetry group to $C_N\times \Z_2^\rF$. Then we combine the construction in [\onlinecite{Yang2024}] with the symmetry-extension method~\cite{Wang2017locWWW1705.06728} arriving at a $\Z_4$ gauge theory as the gapped boundary state. We show that the $C_N$ symmetry in the $\Z_4$ gauge theory is implemented by a static network of invertible topological defects. Interestingly when $\nu = N/2$, we also construct a fully gapped but highly anisotropic state which cannot be described as a TQFT with topological defects.

Let us now discuss some future directions.

First, it will be insightful to understand our results in the context of the decorated domain wall construction. For fermionic SPT phases, the construction produces several ``layers" of decorations~\cite{Wang_2020}. In \spd{4}, for a symmetry group $G$, the decorations are $p_x+\ii p_y$ superconductors on 2D defects (classified by $\H^2(G, \Z)$), Majorana chains on 1D defects (classified by $\H^3(G, \Z_2)$), and complex fermions on 0D defects (classified by $\H^4(G,\Z_2)$). One can show that the $\nu=N$ state is characterized by the Majorana chain decoration, while those with $p_x+\ii p_y$ decorations do not admit a gapped boundary TQFT description. A natural conjecture is that the same is true for the boundary topological orders of \spd{4} fermionic SPT phase with other symmetry groups. Proving or disproving this conjecture is an interesting question for future work. 

 We now give an algebraic argument that the $p+ \ii p$ layer cannot be trivialized by symmetry extension, i.e. the 5-dimensional term
$$A \cup p_1 \equiv p_1(\PD(A)),$$
where $A \in \H^1(G, \U(1))$ paired with the Pontryagin class $p_1$,  
cannot be trivialized by symmetry extension for $G$ being a finite group. Given that all (3+1)d TQFTs are gauged finite group SPTs, this argument strongly suggests that no symmetric boundary TQFT exists for (4+1)d fermionic SPTs with nontrivial $p+\ii p$ decoration.

 The $p+\ii p$ decoration is classified by $\H^2(G, \Z)$. Since $G$ is finite, $\H^2(G, \Z)$ is isomorphic to $\H^1(G, \U(1))$, i.e. group of one-dimensional representations of $G$. Given a nontrivial $\omega \in \H^2(G, \Z)$,  we have a nontrivial 1D representation $\lambda \in \H^1(G, \U(1)) \cong \Hom(G,\U(1))$ ($\omega$ is related to $\lambda$ via the Bockstein homomorphism). Consider a group extension of $G$ by a finite group $K$: 
\begin{equation*}
  1\rightarrow K \rightarrow G'\stackrel{\mathrm{r}}{\rightarrow} G\rightarrow 1, 
\end{equation*} 
where the surjective map from $G'$ to $G$ is denoted by a reduction $\mathrm{r}$. Suppose $\lambda$ is lifted to $\lambda'\in \H^1(G', \U(1))$, then the pullback requires $\lambda'(g')= (\mathrm{r}^* \lambda)(g'):= \lambda(\mathrm{r}(g'))$ for $g'\in G'$. As $\mathrm{r}^*$ is injective on $\H^1(-,\U(1))$, a nontrivial $\lambda$ lifts to a nontrivial $\lambda'$, giving a nontrivial class in $\H^2(G', \Z)$. Therefore, $\omega \in \H^2(G, \Z)$ cannot be trivialized by symmetry extension.

In general, topologically ordered boundary states for FSPT phases with Majorana chain decorations are not well-understood. In \spd{2}, the only known case is the semion-fermion theory, which carries $\nu=2$ anomaly (Majorana chain decoration) of the \spd{3} topological superconductor with time-reversal symmetry. However, for Majorana FSPTs in \spd{3} protected by unitary finite-group symmetry, no explicit examples of topological boundary states are known.
These questions are special cases of a broader problem of characterizing 't Hooft anomalies of topological orders enriched by global symmetries. Namely, given a TQFT and how symmetry acts in the theory, how can we compute the 't Hooft anomaly associated with the symmetry using the given data? While this problem has been extensively studied in \spd{2} (see e.g. \cite{RelativeAnomaly, Tata_2022, BulmashPRB2022b}), the \spd{3} case has been much less explored, which may be a fruitful avenue for future research.

It will also be interesting to explore the connections between the ``TQFT gappability" and the realizability of the symmetry action by a locality-preserving unitary in lattice models. In particular, \sRef{FidkowskiXu} recently showed that if $\nu$ is not a multiple of $2N$, then the symmetry action cannot be implemented by a shallow-depth circuit. For $\nu=2N$ the corresponding FSPT state has complex fermion decorations and can be described by group super-cohomology models. Explicit finite-depth symmetric disentanglers have been constructed in \sRef{Ellison3DFSPT}. \\

\section*{Acknowledgment}

MC is supported by NSF grant DMR-2424315. JW is supported by LIMS and Ben Delo Fellowship. The authors are listed alphabetically.

\textit{Note added}: 
After the completion of this work, \sRef{debray2025buildanomalous31dtopological} and \sRef{wan2025anomalous31dfermionictopological} appeared, reproducing the result of this work using the symmetry extension approach.

\onecolumngrid

\appendix
\section{\spd{4} SPT with \texorpdfstring{$C_N\times\Z_2^\rF$}{CN} symmetry} \label{App:C_N classification}

Here we use the block state construction to derive the fermionic $C_N \times \Z_2^\rF$ SPT classification in \spd{4}. We first consider (3+1)d blocks without any internal symmetry besides $\Z_2^\rF$. Since (3+1)d fermionic SRE phases are all topologically trivial without any symmetries, this layer of decoration is trivial. Then we consider the \spd{2} rotation plane, on which the $C_N$ symmetry becomes an internal $\Z_N$ symmetry. \spd{2} $\Z_N\times\Z_2^\rF$ SRE phases can be divided into two classes:
\begin{enumerate}
    \item Chiral topological superconductors (TSCs) with $c_-=\nu/2$ (i.e. $\nu$ copies of $p_x+\ii p_y$ superconductor);
    \item Fermionic SPT states with no chiral central charge.
\end{enumerate}

Recall that fermionic SPT phases (with $c_-=0$) in \spd{2} with $\Z_N\times\Z_2^\rF$ symmetry has the following classification~\cite{WangPRB2017}:
\begin{equation*}
    \begin{cases}
    \Z_N & \quad N\equiv 1\,(\text{mod }2)\\
        \Z_{4N}=\Z_8\oplus \Z_{N/2} & \quad N\equiv 2\,(\text{mod }4)\\
        \Z_{2N}\oplus\Z_2 & \quad N\equiv 0\,(\text{mod }4)
    \end{cases} \,.
\end{equation*}
This classification can be seen from the domain wall decoration. We list the relevant decorations below
\begin{itemize}
    \item $\mathrm{H}^3(\Z_N, \mathrm{U(1)})$ is the pure bosonic layer;
    \item $\mathrm{H}^2(\Z_N,\Z_2)$ corresponds to complex fermion decoration;
    \item $\mathrm{H}^1(\Z_N,\Z_2)$ corresponds to Majorana chain decoration.
\end{itemize}
When $N$ is odd, the classification is $\Z_N$, consisting of essentially bosonic SPT states. When $\frac{N}{2}$ is odd, the symmetry group factorizes as $\Z_N=\Z_2\oplus \Z_{N/2}$. Thus for $N\equiv 2$ mod 4, the ``fermionic" part comes entirely from the $\Z_2$ subgroup, giving rise to the $\Z_8$ factor. The $N\equiv 0$ mod 4 case is more interesting and we shall return to explicit constructions later.

To physically interpret the \spd{2} fermionic SPT classification, we notice that SPT phases can be physically characterized by the universal properties of the symmetry fluxes. A unit $\Z_N$ flux can be associated with two pieces of universal data. First of all, one can ask whether the flux carries a Majorana zero mode (MZM) or not. We define an invariant $\lambda=0,1$ as the number of MZMs mod 2. Second, the flux has exchange statistics $\theta$, whose $N$-th power $\Theta=\theta^N$ is well-defined. We note that $\theta$ can be related to the spin $h$ of the $\Z_N$ defect operator in the (1+1)d edge CFT of the SPT state via $\theta=e^{2\pi \ii h}$. We can then use these two invariants to characterize the \spd{2} $\Z_N\times\Z_2^\rF$ fermionic SPT phases:

\begin{enumerate}
    \item For $N\equiv 2\,(\text{mod }4)$, $\Theta$ satisfies $\Theta^{4N}=1$, and defines the topological invariant for the phases. The generating phase has $\Theta=e^{\frac{\ii\pi}{2N}}, \lambda=1$. Four copies of the generator has $\Theta=e^{\frac{2\pi \ii}{N}}, \lambda=0$, which is in fact a bosonic state.
    \item For $N\equiv 0\,(\text{mod }4)$, there are two invariants $\Theta$ and $\lambda$. The generator of $\Z_{2N}$ has ${\Theta}=e^{\frac{\pi \ii}{N}}, \lambda=0$, and the generator of $\Z_2$ has ${\Theta}=1$, $\lambda=1$. Notice that two copies of the $\Z_{2N}$ generator state has $\Theta=e^{\frac{2\pi \ii}{N}}, \lambda=0$, which is in fact a bosonic $\Z_N$ SPT state.
\end{enumerate}

When $N \equiv 0$ (mod 4), to further determine the classification of \spd{4} $C_N \times \Z_2^\rF$ fermionic SPT, we need to study the equivalence relations among the block states, labeled by the triplet
\begin{equation*}
    (\nu, m, \lambda), \, \nu\in \Z,\, m\in \Z_{2N},\,\lambda\in \Z_2 \,,
\end{equation*}
where the index $\nu$ labels the $\Z$ classification of $p+ip$, $m$ labels the order of the topological spin of the $\Z_N$ flux and $\lambda$ indicates the existence of MZM. We start with a topologically trivial block state, consisting of $N$ layers of $p+ip$ superconductors, on which the $C_N$ acts as the $\Z_N$ cyclic permutation of the layers. 

To determine the value of $m_N$ for this trivial configuration, we study the edge modes of the state. Each $p+ip$ superconductor contributes a Majorana-Weyl field $\gamma_{L,j}$, $j=1, \cdots, N$. Under $\Z_N$ cyclic layer permutation, they transform as 
\begin{gather*}
    \gamma_{L,j} \rightarrow \gamma_{L, j+1} \,.
\end{gather*}
After Fourier transform, we define the new chiral fields
\begin{gather}
\psi_{L,l} = \frac{1}{\sqrt{N}}\sum_j \omega^{lj}\gamma_{L,j},\;\, \omega=e^{\frac{2\pi \ii}{N}} \,
\end{gather}
transforming under $\Z_N$ as $\psi_{L,l} \rightarrow \omega^l \psi_{L,l}$. Since $\psi_{L,l}^\dagger = \psi_{L,N/2-l}$, we identify $\psi_{L,\alpha},\,\alpha = 1,\cdots,N/2$  as independent Weyl fields, and $\psi_{L,0}$, $\psi_{L,N/2}$ Majorana-Weyl fields. Under $\Z_N$ symmetry, $\psi_{L,0}$ remains invariant while $\psi_{L,N/2}$ obtains a minus sign. Therefore, the topological spin of a unit $\Z_N$ defect consists of those from $\frac{N}{2}-1$ Weyl fields and one Majorana-Weyl field $\psi_{L,N/2}$:
\begin{equation}
    h=\frac12 \sum_{l=1}^{N/2-1}\left(\frac{l}{N}\right)^2+\frac{1}{16}=\frac{N^2+2}{48N}\,,
\end{equation}
i.e. $Nh=(N^2+2)/48$. In addition, the unit $\Z_N$ defect carries a Majorana zero mode from $\psi_{N/2}$.

In the following we assume $N=2^p$. $m_N$ can be determined by:
\begin{equation}
e^{\frac{\ii\pi m}{N}}={\Theta}=e^{\frac{\ii\pi}{24}(N^2+2)} \,\; \Longrightarrow \;\,  m_N=\frac{N(N^2+2)}{24}=2^{p-2}\frac{2^{2p-1}+1}{3}.
\end{equation}

The trivial block state corresponds to $(N, m_N,1)$,  thus we find that $(-N,0,0)\sim (0,m_N,1)$ \footnote{$\sim$ means phase equivalence.}. Denote $r_N=\frac{2^{2p-1}+1}{3}$, which is easily seen to be an odd integer. This indicates that $(0, m_N,0)$ has order $8$ in $\Z_{2N}=\Z_{2^{p+1}}$. Therefore, the element $(1,0,0)$ has order $8N=2^{p+3}$. 

On the other hand, $(-2N,0,0)\sim (0,2m_N,2 \equiv 0)$, we have $(-2^{p+1}, 0, 0)\sim (0, 2^{p-1}r_N, 0)$, thus $2^{p-1}\times (4,r_N,0)\sim (2^{p+1},2^{p-1}r_N,0)\sim (0,0,0)$. Since $r_N$ is odd, we conclude that $(1,0,0)$ and $(4,r_N,0)$ generate all the phases. The group structure is thus $\Z_{2^{p+3}}\oplus\Z_{2^{p-1}}$.

 When $N=2$, the decoration is represented by $(\nu, m)$ where $\nu\in\Z, m\in \Z/8\Z$. The topological invariant satisfies $e^{\frac{\ii\pi}{4}m}=\Theta=e^{\frac{\ii\pi}{4}}$, thus $m=1$. We have an equivalence relation $(-2,0,0)\sim (0,1)$. In this case $(1,0,0)$ generates the whole group $\Z_{16}$.

 We conclude that the $C_N \times \Z_2^\rF$ SPT classification in \spd{4} is given as follows:
\begin{itemize}
    \item For $N\equiv 1\,(\text{mod }2)$, the classification is the same as the classification of \spd{4} bosonic $C_N$ SPT, which is derived using the block construction in [\onlinecite{Yang2024}]:
    \begin{equation*}
        \begin{cases}
            \Z_{N}\oplus \Z_N & 3 \nmid N\\
            \Z_{3N} \oplus \Z_{N/3} & 3 \;|\; N
        \end{cases} \, ;
    \end{equation*}
    \item For $N$ even, write $N=q \cdot 2^p$, where $q$ is odd. Then the symmetry group $\Z_{2N}^\rF$ is isomorphic to $\Z_{2^{p+1}}^\rF\times \Z_{q}$. Since $N/2$ is the odd, the classification is the direct sum of $\Z_{2^{p+3}}\oplus \Z_{2^{p-1}}$ (from $\Z_{2^{p+1}}^\rF$) and the one for $\Z_{q}$ in the previous case.
\end{itemize}

\section{\texorpdfstring{$\Z_{16}$}{Z16} class anomaly of \texorpdfstring{$\Z_{4}^{\rm F}$}{Z4} symmetry in the Standard Model}
\label{sec:BSM}

The conventional Standard Model (SM) has the following features:

\begin{enumerate}

\item  {\bf $15\times 3$ Weyl fermions and an anomaly index $-3 + n_{\nu_R}$}:

The SM has $15$ Weyl fermions per family (or per generation), and with $N_f=3$ families thus a total $15\times 3=45$ Weyl fermions of quarks and leptons.
We do not yet know how many types of right-handed neutrinos there are beyond the SM:
the total number of types of right-handed neutrinos, say $n_{\nu_R}$, can be $n_{\nu_R}=0,1,2,3,\dots$ and so on.
The potentially nonzero number $-N_f+n_{\nu_R}=-3 + n_{\nu_R}$ is an important anomaly index for the SM.

\item  {\bf $\U(1)_{{\bf B}- {\bf L}}$ symmetry and anomaly}:

A continuous baryon {\bf B} minus lepton {\bf L} number symmetry, $\U(1)_{{\bf B}- {\bf L}}$ is preserved within the SM.
More precisely, to have a properly quantized charge, it is better to normalize $\U(1)_{{\bf B}- {\bf L}}$ 
by a factor of $N_c$ as
the quark {\bf Q} number minus $N_c$ lepton {\bf L} number symmetry,
$\U(1)_{{\bf Q}- N_c {\bf L}}$, where the color number is $N_c=3$ in the SM.\footnote{Namely,
when we mention {$\U(1)_{{\bf B}- {\bf L}}$ symmetry and anomaly}, we really mean
$\U(1)_{{\bf Q}- N_c {\bf L}}$ symmetry and anomaly.}
But the $\U(1)_{{\bf Q}- N_c {\bf L}}$ symmetry has a 
't Hooft anomaly in 4d spacetime captured by 
a 5d invertible topological field theory (iTFT) in one extra dimension
with the following invertible U(1) functional:
\begin{equation}\label{SM-U1-iTFT-1}
   \exp(\ii  S_5)
\equiv \exp\Bigg[\ii   \int_{M^5} (-N_f+n_{\nu_R}) \,A \wedge \left(N_c^3 \frac{1}{6} \dd A \wedge  \dd A +N_c\frac{1}{24}  \frac{1}{8 \pi^2}   \Tr[ R \wedge R] \right) \Bigg], 
\end{equation}

where $A$ is the U(1) gauge field connection (locally a 1-form) and $R$ is the spacetime curvature locally a 2-form.
In terms of the relation between \eq{SM-U1-iTFT-1} and the perturbative Feynman diagram as triangle cubic term diagrams in \Fig{fig:Feynman}: 

\begin{figure}[!h]
    \centering
\includegraphics[width=.25\columnwidth]{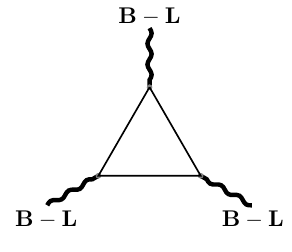}
\includegraphics[width=.25\columnwidth]{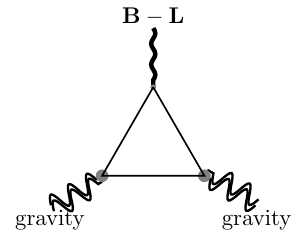}
    \caption{The Standard Model has $\U(1)^3$ anomaly and $\U(1)$-gravity$^2$ anomaly
    with the $\U(1) \equiv \U(1)_{{\bf B}- {\bf L}}$ or more precisely as $\U(1) \equiv \U(1)_{{\bf Q}- N_c {\bf L}}$, captured by two kinds of one-loop triangle Feynman diagrams shown here.
    The anomaly coefficient is given by $(-N_f+n_{\nu_R})$, the integer-value difference between the family number $N_f=3$ and 
    the total number $n_{\nu_R}$ of types of right-handed neutrinos.}
    \label{fig:Feynman}
\end{figure}

the $A \wedge \dd A \wedge  \dd A$ matches with the $\U(1)^3$ anomaly
and the $A \wedge  \Tr[ R \wedge R]$ matches with the $\U(1)$-gravity$^2$ anomaly.
Or more precisely, \eq{SM-U1-iTFT-1} can be written as in terms of Chern and Pontryagin characteristic classes:
\begin{equation}
 \label{SM-U1-iTFT-2}
\exp(\ii  S_5)
\equiv \exp\Bigg[\ii   \int_{M^5} (-N_f+n_{\nu_R}) \,A_{{\bf Q}-N_c {\bf L}}\left(N_c^3 \frac{c_1(\U(1)_{{{\bf Q}-N_c {\bf L}}})^2}{6}-N_c\frac{p_1(TM)}{24}\right) \Bigg] .
\end{equation}
{Here the $n$-th Chern class is denoted as $c_n$, locally $c_1 \coloneqq \int_{M^2} \frac{1}{2 \pi} \dd A$;  
the first Pontrygain class is $p_1$, locally $p_1  \coloneqq \int_{M^4} - \frac{1}{8 \pi^2}   \Tr[ R \wedge R]$. To obtain a nontrivial characteristic class, it requires taking care of the transition functions between local patches glued together to do a global integral on the whole manifold.}

\item  {\bf $\Z_4^\rF = \Z_{4, X \equiv 5({\bf B} - {\bf L})-\frac{2}{3} \tilde{Y}}$ and $\U(1)_X$  
symmetry and anomaly}:

A discrete baryon {\bf B} minus lepton {\bf L} symmetry linear combined with a properly  quantized electroweak hypercharge $\tilde{Y}$,
known as $X \equiv 5({\bf B} - {\bf L})-\frac{2}{3} \tilde{Y}
= \frac{5}{N_c}({{\bf Q}- N_c {\bf L}})-\frac{2}{3} \tilde{Y}$ is identified by 
Wilczek-Zee, which can be a $\U(1)_X$ preserved in the SM. 
But the $\U(1)_X$ symmetry has the same 't Hooft anomaly captured by \eq{SM-U1-iTFT-2}.

This specific combination as an order four finite cyclic group $\Z_{4, X \equiv 5({\bf B} - {\bf L})-\frac{2}{3} \tilde{Y}}$ is also pointed out by Wilczek-Zee~\cite{Krauss:1988zcWilczek}.
The $\Z_{4, X}$ symmetry in the SM is exactly the $\Z_{4}^{\rm F}$ symmetry
discussed in our main text.
Recently, in [\onlinecite{Garc_a_Etxebarria_2019, Wan_2020,
wang2020anomalycobordismconstraintsstandard, wang2020anomalycobordismconstraintsgrand, JW2012.15860,
WangWanYou2112.14765, WangWanYou2204.08393, Putrov_2024}],
it was pointed out that the (3+1)d SM suffers from a mod 16 class of this mixed $\Z_{4, X}$-gauge-gravity nonperturbative global anomaly captured by a (4+1)d fermionic invertible topological field theory
(namely, the low-energy field theory of the (4+1)d FSPT that we studied)
\bea  \label{SM-Z16-iTFT} 
\exp(\ii S_5)  
\equiv \exp\Bigg[\ii  (-N_f+n_{\nu_R})\,
     \frac{2\pi }{16}  \eta_{4{\dd}}(\text{PD}( A_{{\Z_{2,X}}} )) \big\vert_{M^5}\Bigg].
\eea
The background gauge field $A_{{\Z_{2,X}}}\in \H^1(M^5,\Z_2)$ is 
obtained by the quotient map $\Z_{2,X}\equiv {\Z_{4,X}}/{\Z_2^\rF}$ from 
the ${\Spin \times_{\Z_2^\rF} {\Z_{4,X}}}$-structure on the 5d spacetime manifold $M^5$.
The 4d Atiyah-Patodi-Singer eta invariant $\eta_{4{\dd}}$ 
is the $\Z_{16}$ class of topological invariant of time-reversal symmetric topological superconductor (with a time-reversal generator $\rm{T}$ whose
$\rm{T}^2=(-1)^{\rm F}$).
The $\eta_{4{\dd}}$ is evaluated at the 4d submanifold Poincar\'e dual (PD) to the $A_{{\Z_{2,X}}}$ in the 5d bulk.

{For the quarks and leptons in the SM and the right-handed neutrinos,
we can write them all 
as a multiplet of the left-handed Weyl fermions 
(as two-component Lorentz spinors) under the 
Lie algebra representation of
$su(3) \times  su(2) \times u(1)_{\tilde Y} \times u(1)_X$
\begin{equation}\label{eq:SMrep}
  \begin{aligned}
({\psi_L})_{\rm I} =
( \bar{d}_R \oplus {l}_L  &\oplus q_L  \oplus \bar{u}_R \oplus   \bar{e}_R  
)_{\rm I}
\oplus
n_{\nu_{{\rm I},R}} {\bar{\nu}_{{\rm I},R}}
\\
&\sim 
\big((\overline{\bf 3},{\bf 1})_{2,-3} \oplus ({\bf 1},{\bf 2})_{-3,-3}  
\oplus
({\bf 3},{\bf 2})_{1,1} \oplus (\overline{\bf 3},{\bf 1})_{-4,1} \oplus ({\bf 1},{\bf 1})_{6,1} \big)_{\rm I}
\oplus n_{\nu_{{\rm I},R}} {({\bf 1},{\bf 1})_{0,5}}. 
\end{aligned}   
\end{equation}

The family index is ${\rm I}=1,2,3$
with ${\psi_L}_1$ for $u,d,e$ type,
${\psi_L}_2$ for $c,s,\mu$ type,
and 
${\psi_L}_3$ for $t,b,\tau$ type of quarks and leptons.
We use ${\rm I}=1,2,3$ for $n_{\nu_{e,R}}, n_{\nu_{\mu,R}}, n_{\nu_{\tau,R}} \in \{ 0, 1\}$
to label either the absence or presence of electron $e$, muon $\mu$, or tauon $\tau$ types of sterile neutrinos (i.e., ``right-handed'' neutrinos sterile to 
$\cG_{\rm SM} \equiv su(3) \times  su(2) \times u(1)_{\tilde Y} $ gauge forces).
Here 
the total number of types of right-handed neutrinos is $n_{\nu_R} \equiv 
n_{\nu_{e,R}} + n_{\nu_{\mu,R}} + n_{\nu_{\tau,R}}$.\\
The SM multiplet has 
$({\psi_L})_{\rm I} \sim \bar{\bf 5}_{-3} \oplus {\bf 10}_1 \oplus {\bf 1}_5$
under $su(5) \times u(1)_X$ for SU(5) grand unification, while 
$({\psi_L})_{\rm I} \sim  {\bf 16}_1$
under $so(10) \times \Z_{4,X}$ for Spin(10) grand unification.
}

All quarks and leptons 
in the SM has a charge 1 under $\Z_{4, X}$.
Furthermore, the beyond-the-SM (BSM) right-handed neutrino (here written as the left-handed Weyl fermion particle) $\bar{\nu}_R$ also has $\Z_{4, X}$ charge 1.
Thus, the scaling dimension-6 (in energy) 
four-fermion deformation operator has a charge $4 = 0 \mod 4$ under $\Z_{4, X}$. 
The $\Z_{4, X}$ symmetry is still preserved in the modified SM with these BSM deformations:

\begin{equation*}
\psi_q\psi_q\psi_q\psi_l, \quad \psi_{\bar q}\psi_q\psi_{\bar l}\psi_l, \quad 
\psi_{\bar q}\psi_q \psi_{\bar q}\psi_q, \quad \psi_{\bar l}\psi_l\psi_{\bar l}\psi_l.    
\end{equation*}

The $\psi_q$ and $\psi_l$ are the quark and lepton fermion field operators from the SM multiplet in \eq{eq:SMrep}.
Some of these four-fermion BSM deformations violate 
the baryon {\bf B} conservation (which triggers nucleon decays)
or lepton {\bf L} conservation, or ${\bf B} - {\bf L}$ 
conservation \cite{PhysRevLett.43.1566, Wilczek1979hcZee}.
Some of these four-fermion BSM deformations can preserve the full $\U(1)_X$, 
but some of these BSM deformations can break $\U(1)_X$ down to $\Z_{4, X}$.

As mentioned in the introduction, the $\Z_{16}$ class of the $\Z_4^\rF$ chiral anomaly shows up as the mixed gauge-gravity nonperturbative global anomaly for $\Z_{4, X}$ in the SM. Three generations of experimentally confirmed quarks and leptons contribute a total number of Weyl fermions in the SM $3 \times 15 = 45$, which gives an anomaly index $45 \mod 16 = -3 \mod 16$. The $-3 \mod 16$ is the anomaly index of the SM, for the $\Z_{16}$ class of the $\Z_{4,X}$-gauge-gravitational nonperturbative global anomaly. As a speculation,
{the (3+1)d boundary topological order that we construct together with the (4+1)d bulk FSPT provides an alternative way to cancel the $-3 \mod 16$ class of anomaly in the SM, without necessarily including 3 generations of right-handed neutrinos. The experimental absence of any right-handed neutrinos in the SM provides us the theoretical opportunity to include the exotic topological BSM sector via appending the (3+1)d non-invertible topological order or (4+1)d invertible topological phase to the SM. For example, anomaly cancellation can be achieved by having one right-handed neutrino, together with the \spd{3} $\Z_4$ gauge theory TQFT discussed here. Thus, these topological states (and the fractionalized/anyonic energetic excitations above the topological order energy gap) become a candidate for the BSM cold dark matter~\cite{wang2020anomalycobordismconstraintsstandard, wang2020anomalycobordismconstraintsgrand, JW2012.15860}.}

\end{enumerate}
 
\twocolumngrid

\bibliography{fermion-bdry}

\end{document}